# Study of low frequency acoustic signals from superheated droplet detector

**Prasanna Kumar Mondal, Susnata Seth, Mala Das and Pijushpani Bhattacharjee**

Astroparticle Physics & Cosmology Division, Saha Institute of Nuclear Physics,
1/AF Bidhannagar, Kolkata 700064, India

E-mail: prasanna.mondal@saha.ac.in, susnata.seth@saha.ac.in, mala.das@saha.ac.in and pijush.bhattacharjee@saha.ac.in

**Abstract**

The bubble nucleation process in superheated droplet detector (SDD) is associated with the emission of an acoustic pulse that can be detected by an acoustic sensor. We have studied the neutron and gamma-ray induced nucleation events in a SDD with the active liquid R-12 ($CCl_2F_2$, b.p. -29.8$^o$C) using a condenser microphone sensor. A comparative study in the low frequency region (~ 0-10kHz) for the neutron and gamma-ray induced nucleation is presented here. From the analysis of the waveforms we observe a significant difference between the neutron and gamma-ray induced acoustic events.

# 1. Introduction

A superheated droplet detector (SDD) [1, 2] consists of a large number of micron sized superheated liquid droplets dispersed in a viscoelastic gel [1] or in a soft polymer [2] medium. SDD is known to be able to detect neutrons [3, 4], gamma-rays [5, 6] and other charged particles [7, 8] under certain operating temperature and pressure [5, 9]. In SDD, each droplet can be considered as a micron size bubble chamber [10-12], where a drop vaporizes when sufficient energy is deposited by the energetic radiation. The vaporization of a superheated droplet is associated with a change in volume and also the emission of an acoustic pulse, both of which can be detected independently by active devices [13-16]. The acoustic pulse bears some unique characteristic features, which makes SDD a good tool for discrimination of different kinds of energetic radiation [17, 18]. In recent years, the study of the characteristics of the acoustic signals is one of the important topics in the superheated liquid based detector research [16-20].

SDD is widely used in neutron spectrometry [21-24] and neutron dosimetry [23-26], and considered as a efficient neutron detector in specified temperature and pressure regions, where it is not sensitive to gamma-rays [4, 22, 27, 28]. The gamma-ray sensitive temperature provides a lower cut on the neutron detection in a mixed neutron-gamma ray field. The separation of neutron and gamma-ray is important at high temperature region when the recoils have lower LET comparable with the energy deposition by the electrons. Therefore, for the detection and dosimetry of neutrons, in presence of gamma-ray background, the neutron-gamma ray discrimination becomes important. Another importance of the neutron and gamma-ray discrimination is in the dark matter search experiments using superheated liquid based detectors [12, 29, 30]. Superheated liquid or droplet detector is widely used for the dark matter search specially WIMPs (weakly interacting massive particles) by different groups (COUPP [12], PICASSO [29], SIMPLE [30]). WIMPs undergo elastic scattering with the detector nuclei and produce the recoil nuclei with energy in the range of 10-100 keV. These recoil nuclei then deposit the energy similar to the neutron induced recoil nuclei and subsequently produce the bubble if bubble formation condition is satisfied. As the WIMP induced nuclear recoils are similar to the neutron induced recoils, the WIMP signal is expected to be in the temperature region of the neutron signals. The importance of separating the gamma-ray signal from the dark matter signal is the same as that of the neutron. Several groups (DAMA/LIBRA [31], CoGeNT [32], CRESST [33]) have claimed possible dark matter signal corresponding to WIMPs mass below 10 GeV/$c^2$. In order to observe this effect, detectors with very low recoil energy threshold and good separation from different types of backgrounds including gamma-rays are required.

In SDD, the detection of the acoustic pulse is made by the acoustic sensors like piezoelectric transducers [13] or by microphones [16, 34]. In this work, the acoustic pulses from the bubble nucleation in SDD are detected by using a condenser microphone sensor [34]. The microphone converts the acoustic pulse into an electrical signal. The aim of this work is to study various parameters of the acoustic signals for neutron and gamma-ray induced events and to find out the parameters that can be effectively used for neutron-gamma ray discrimination. We have studied the lower frequency region of the signal (up to 10 kHz), within which the principal frequency components of the signal are generally found [16, 20]. We have used a superheated R-12 ($CCl_2F_2$, b.p. -29.8°C) based SDD where the bubble nucleation was induced by using $^{241}$Am and $^{137}$Cs gamma-ray sources and a gamma-ray shielded $^{241}$AmBe neutron source. Different parameters of the microphone signals are studied to discriminate the neutron and gamma-ray induced events in SDD. Earlier, Das *et. al.* [18] have studied the maximum amplitude and signal power with R114 ($C_2Cl_2F_4$; b.p. 3.7°C) liquid based SDD using $^{252}$Cf neutron source and $^{137}$Cs gamma-ray source. In this work, the parameters like the maximum pulse height ($V_{max}$) of the waveform, total signal power ($P$), principal harmonic frequency ($F$), integral area under the principal harmonic frequency peak ($P_F$) of the power spectral density spectra and the time constant ($\tau$) of the waveform, are studied.

From this study it is found that the time constant parameter distribution for the neutron and gamma-ray induced events are difficult to distinguish from each other, whereas the other parameter distributions show significant differences for neutron and gamma-ray induced events, which can be useful for neutron-gamma ray discrimination in a mixed neutron-gamma ray radiation field.

## 2. Principle of Measurement

The bubble nucleation process in superheated liquid is described by the Seitz's thermal spike model [35]. According to this model, as the energetic radiation deposits energy along its path inside the superheated liquid, embryonic vapour bubbles or microbubbles are formed along the path. If a microbubble has a size larger than a critical size ($r_c$), the bubble nucleation occurs, and only then the vapour bubble can grow spontaneously to observable size by vaporizing the superheated liquid. For bubble nucleation to occur, the energy deposition must be greater than the minimum energy ($W$) needed to form a critical size microbubble. Thus $W$ is the threshold energy for bubble nucleation and is a function of temperature, pressure and the liquid-vapour interfacial surface tension. The bubble nucleation and the liquid-vapour phase transition in superheated liquid is a fast process (occurs on a time scale of few micro seconds) during which an acoustic shock wave is produced. After complete vaporization of a superheated droplet, the resulting bubble oscillates freely around its equilibrium radius with a resonance frequency [16, 36]. Along with this resonance frequency some higher frequency components are also observed when a piezoelectric transducer is used for the detection of acoustic signals, whereas the microphone detects only the low frequency part of the signal.

## 3. Experimental Method

The SDD used in this work is prepared by a simple emulsification process [37] using superheated liquid droplets of R-12 dispersed in a viscoelastic gel matrix. The perspex container holding the gel and the liquid drops is wrapped with a heating coil connected to a temperature controller and temperature sensor. The experimental setup is shown in figure 1. Here the condenser microphone sensor was placed in such a way that its active surface touches the upper surface of the emulsion. The condenser microphone contains two parallel plates one of which is fixed and the other is movable (called the diaphragm) with a small gap between them. When an acoustic signal falls on the diaphragm, the distance between the plates changes and as a result the capacitance between the plates changes which gives rise to the electrical signal. The output from the microphone circuitry was fed directly to an oscilloscope (Agilent Technologies, MSO7032A) without any external amplification. The electrical signals from the microphone were acquired using the oscilloscope operating in pulse acquire mode with a sampling frequency of 50 kS/s.

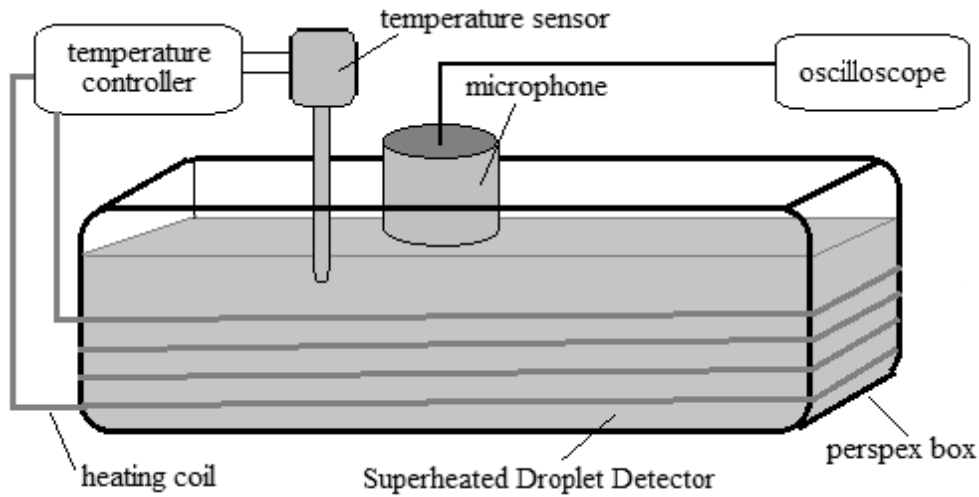

**Figure 1**. Experimental setup.

To study the neutron induced nucleation, the emulsion was irradiated with a gamma-ray shielded 3 Ci $^{241}$AmBe neutron source at the average detector temperatures of about 40.5°C and 44.5°C and the pulse shape data were recorded in ASCII file. For the gamma-ray shielding of the neutron source, about 8 cm thick lead bricks were used. The gamma-ray induced nucleation was studied using 0.5 Ci $^{241}$Am and 0.02 Ci $^{137}$Cs gamma-ray sources. It is observed earlier that for $^{241}$Am and $^{137}$Cs gamma-ray sources the gamma-ray sensitive threshold temperature for R-12 is about 38.5°C and 42.5°C respectively [5-6]. Since R-12 detector is insensitive to $^{137}$Cs gamma-ray source at 40.5°C the pulse shape data for $^{137}$Cs gamma-ray source were acquired at about 44.5°C and for $^{241}$Am gamma-ray source the data were acquired at about 40.5°C and 44.5°C. The detector sensitivity in terms of the number of events per 10 µSv neutron and gamma-ray dose at 40.5°C and 44.5°C are shown in figure 2. The neutron dose rate is measured using neutron monitor NM2B (NE Technology Ltd, UK) and the gamma-ray dose rate is measured using PM1402M radiation monitor (Polimaster Ltd, Republic of Belarus). The data files in ASCII were converted, using a FORTRAN code, to NI LabVIEW measurement files for plotting and further analysis. The typical waveforms of the neutron and gamma-ray induced events at 40.5°C are shown in figure 3. From the traces of the pulses, various parameters have been analyzed, as explained in next section.

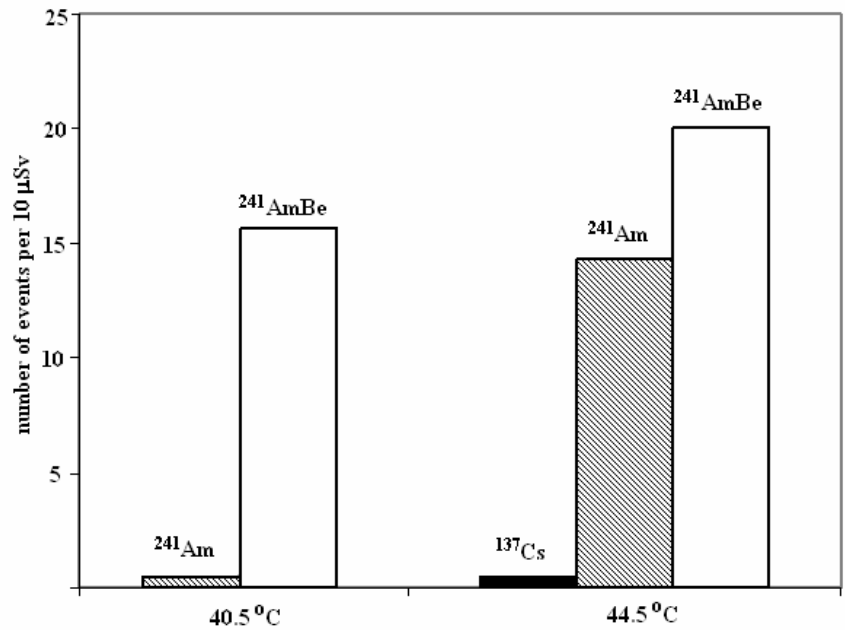

**Figure 2**. Neutron and gamma-ray dose response of the detector for $^{241}$AmBe neutron source and $^{241}$Am and $^{137}$Cs gamma-ray sources. Here the bars represent the number of bubble nucleation events per 10 µSv dose.

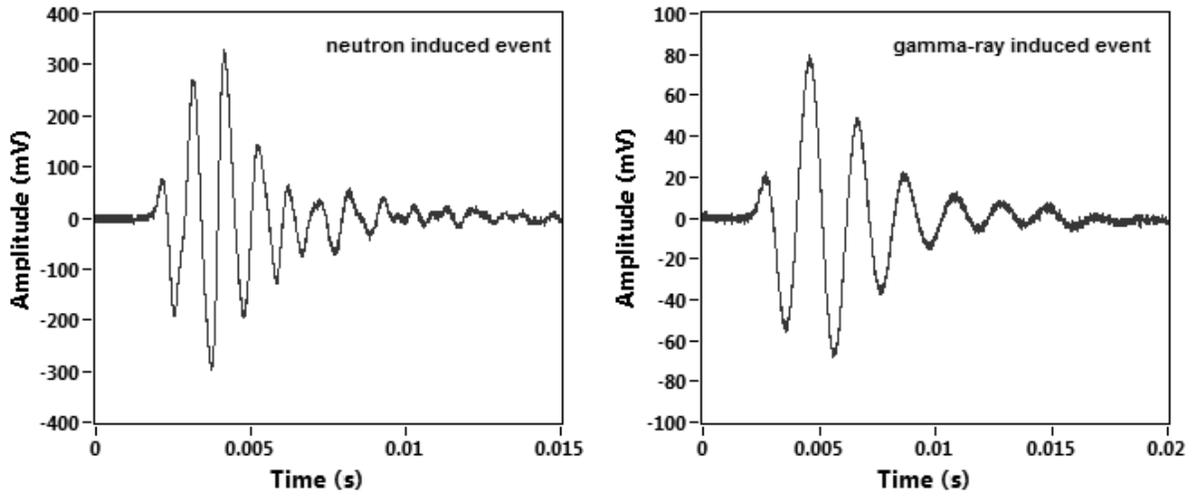

**Figure 3**. Typical waveform of the condenser microphone output for neutron and gamma-ray induced bubble nucleation events.

## 4. Results and Discussion

In this work, the traces of the pulses are analyzed using LabVIEW program. A high pass filter of 150 Hz is used in LabVIEW program to reduce the electronic noise level in the acquired data. The filtered waveforms are then used in the next step of the analysis as explained below, in order to explore the different parameters, useful for discrimination between neutron and gamma-ray induced events.

### 4.1. Measurement of the time constant of the signal

A typical signal waveform shown in figure 3 usually has a fast rise at the beginning and then falls slowly as damped oscillations. The decay time constant analysis of the signal is obtained with the amplitude envelope of the waveform. The amplitude envelope of the waveform is calculated using the Hilbert transform and a typical envelope of the pulse is shown in figure 4(a). The decaying part of the envelope is fitted to an exponential function $y(t) = Ae^{-t/\tau}$, and from the best fitted curve (figure 4(b)) the decay time constant $\tau$ is obtained. The distribution of $\tau$ for neutron and gamma-ray induced events at 40.5°C and 44.5°C are shown in figure 5. The result indicates that the time constant distribution of the gamma-ray induced signals strongly overlaps with that of the neutron induced events, and is therefore, not useful for the discrimination of neutron and gamma-ray induced events.

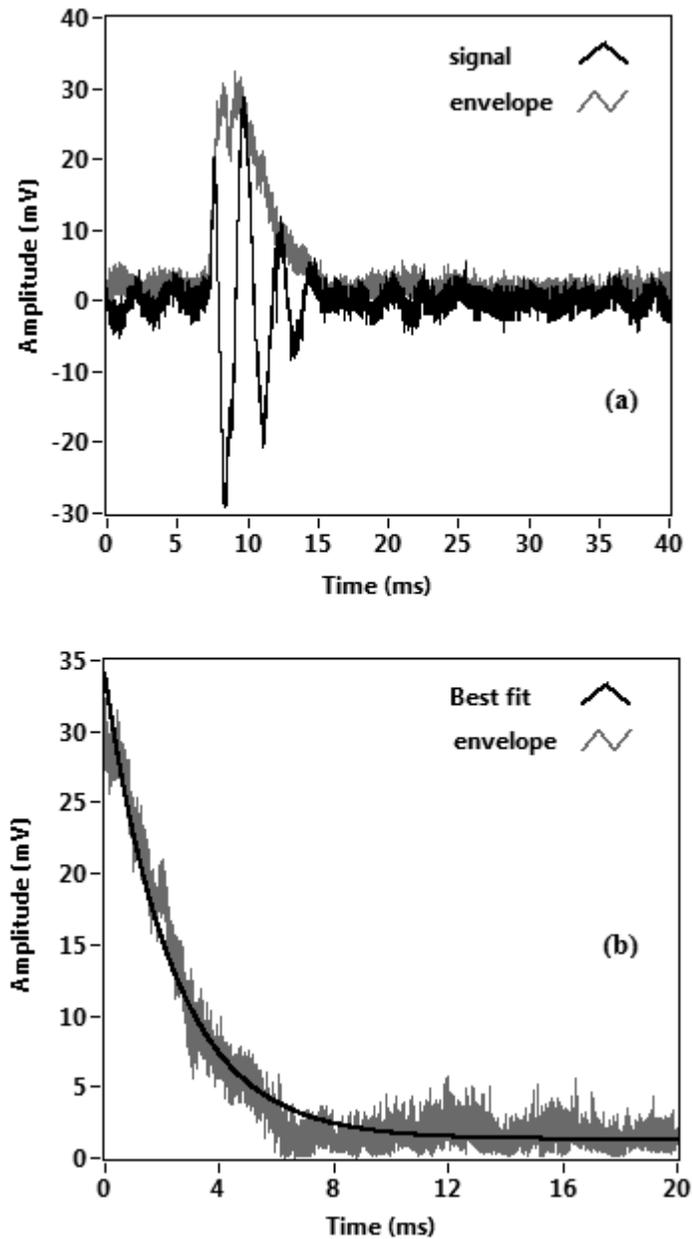

**Figure 4**. A typical envelope of a waveform as obtained using Hilbert transform (a) and the exponential fit of the decaying part of the envelope (b).

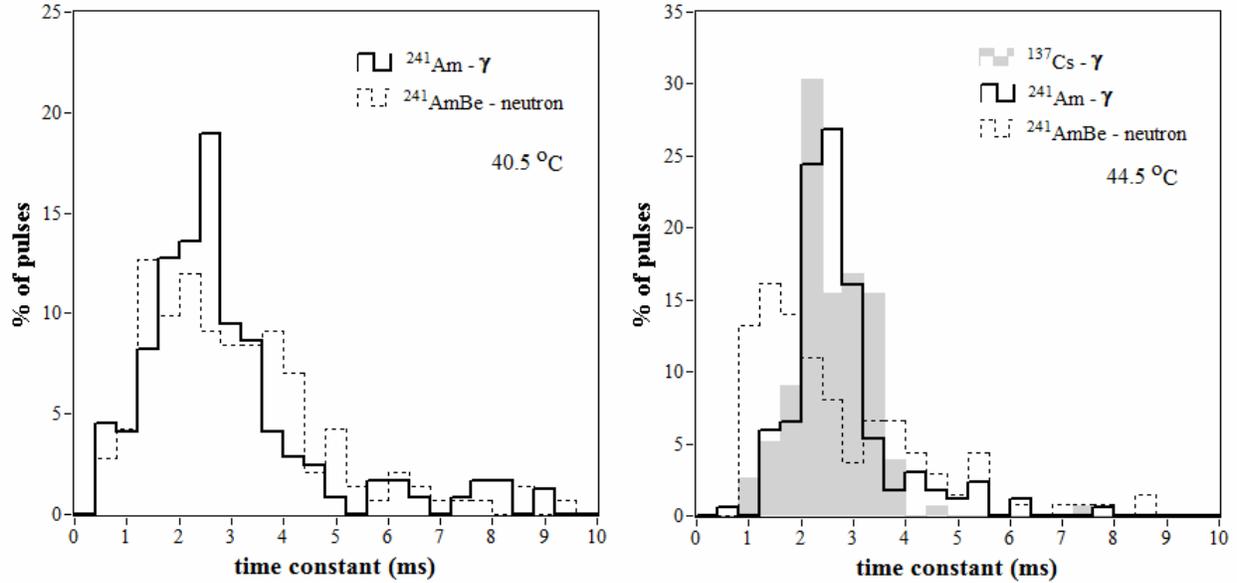

**Figure 5**. Distribution of the decay time constant of the pulses obtained with microphone for the neutron and gamma-ray induced nucleation events at 40.5°C and 44.5°C.

### 4.2. Measurement of maximum pulse amplitude

The maximum amplitude ($V_{max}$) of each pulse is obtained for neutron and gamma-ray induced events. The amplitude distribution for neutron induced events is compared with those of the gamma-ray induced events at 40.5°C and 44.5°C (figure 6). Figure 6 shows that the pulses due to the neutron induced events are predominantly of higher amplitude compare to those of the gamma-ray induced events, which agrees with the earlier report [18] by Das *et. al*. This happens due to the fact that for neutrons the heavy recoils, with their higher LET, deposit more energy in the critical diameter compared to that deposited for gamma-rays, since the electrons deposit energy almost at the end of their track [23].

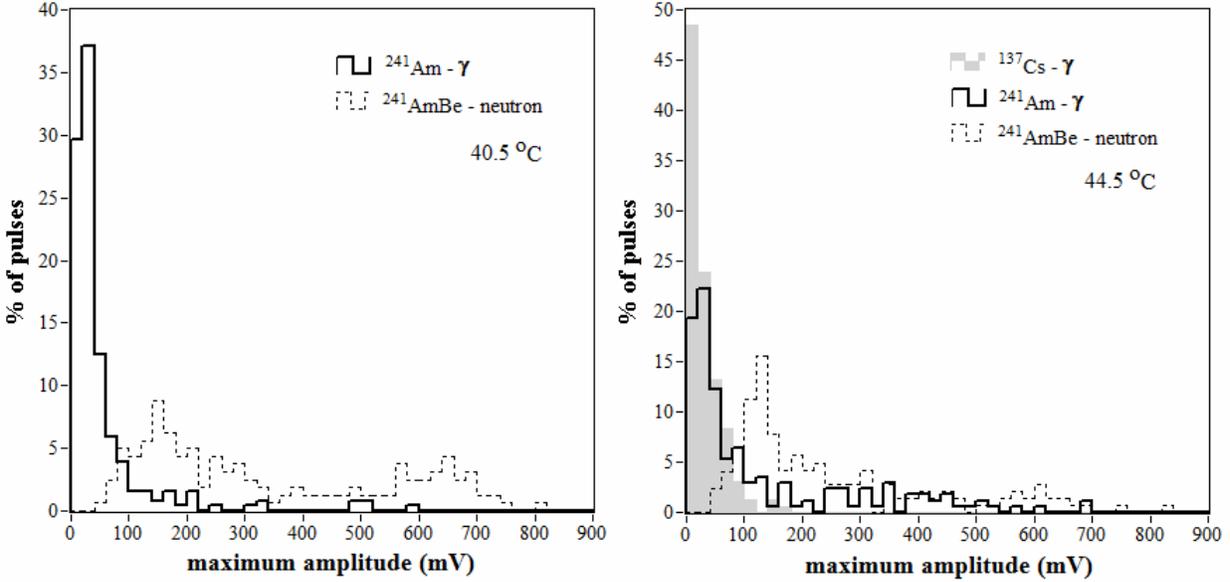

**Figure 6**. The maximum amplitude distribution of the waveforms obtained for neutron and gamma-ray induced events at 40.5°C and 44.5°C.

### 4.3. Measurement of signal power using the pulse waveform

The maximum amplitude of the waveform provides a rough estimate of the energy deposited and the energy released during the nucleation event. The power of the signal is proportional to the energy released during the bubble formation process and thus important information on the released energy can be obtained from the power distribution of the signal. It is a measure of the energy contained in the microphone signal. Here the power, $P$, is defined as $P = \ln\left(\sum_i V_i^2\right)$, where $V_i$ is the signal amplitude at the $i^{th}$ time bin and the summation extends over the duration of the signal. In figure 7, the histograms for the power variations of the neutron and gamma-ray induced events at 40.5°C and 44.5°C are shown, where the $P$-axis is rescaled to $P = 0$ corresponding to the lowest $P$ value obtained. It shows that the gamma-ray induced events predominantly are of lower $P$ values than those of the neutron induced events. This implies that in the neutron induced events more acoustic energy is emitted compared to that emitted in the gamma-ray induced events.

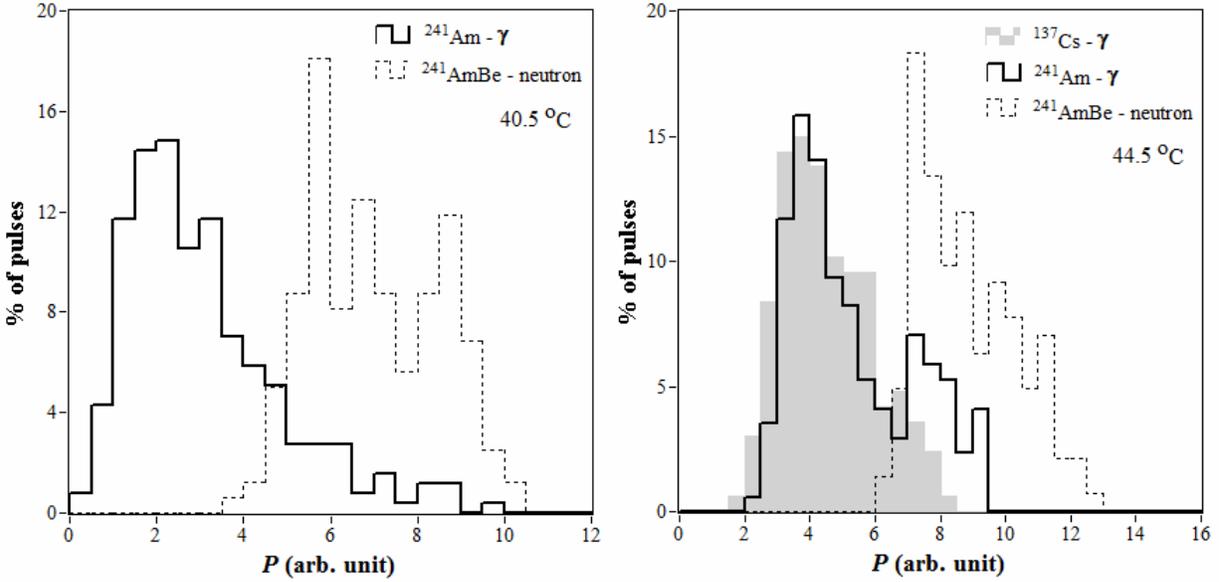

**Figure 7**. The power distribution of the pulses obtained from the waveform for neutron and gamma-ray induced events at 40.5°C and 44.5°C.

### 4.4. Measurement of frequency of the primary harmonic

The spectral density or the power spectral density, PSD, describes how the power of a signal is distributed as a function of the frequency. From the PSD distribution of the waveform, we have obtained the frequency of the primary harmonic $F$, defined as the frequency corresponding to the highest peak in the spectral density distribution of the signal. The distribution of $F$ for the neutron and gamma-ray induced nucleation at 40.5°C and 44.5°C are shown in figure 8. From figure 8 it is observed that the neutron and gamma-ray induced events are well separated in the $F$ distribution plots. With increase in temperature, the $F$ distribution shifts towards higher frequency region and the separation between the $F$ distributions for the neutron and gamma-ray induced events decreases. A gamma-ray induced nucleation event at 40.5°C possesses a characteristic frequency response, with a primary harmonic between 200–500 Hz. Here at 40.5°C about 87% of gamma-ray induced events have $F$ parameter values below 500 Hz, while about 97% of the neutron induced events have $F$ values above 500 Hz. Thus the condenser microphone sensor provides good discrimination between the neutron and gamma-ray induced events from the analysis of primary harmonic parameter.

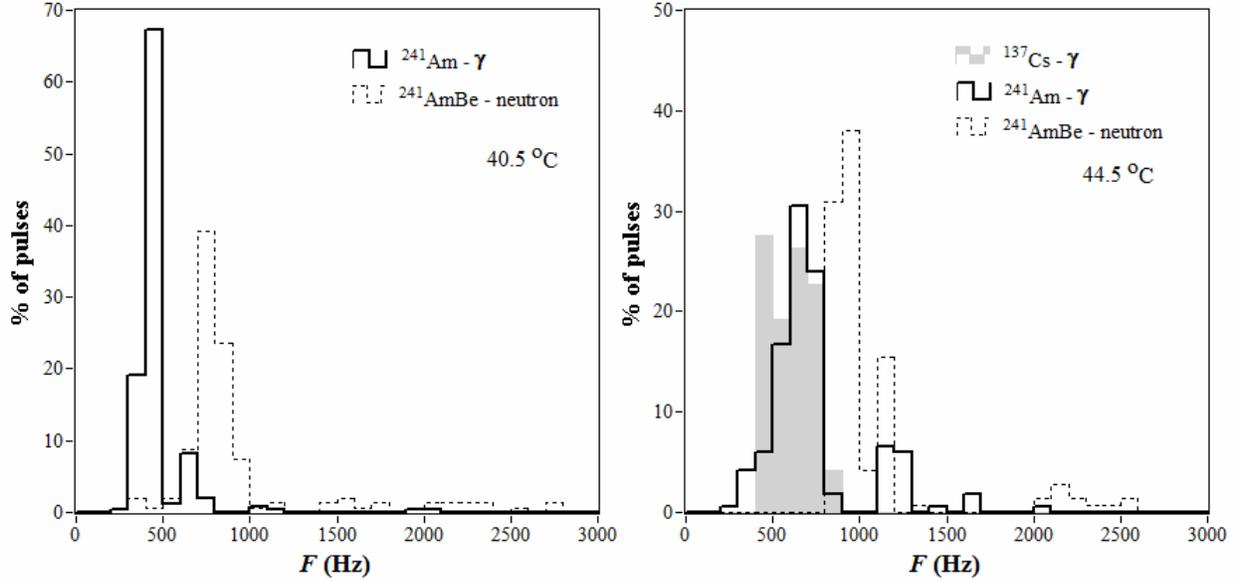

**Figure 8**. Distribution of the frequency of the primary harmonics of the pulses for neutron and gamma-ray induced events at 40.5°C and 44.5°C.

## 4.5. Measurement of signal power about the primary harmonic (*F*) of the waveform

The power spectral density, PSD, gives a measure of the signal power. From PSD distribution of the signal, we have obtained the frequency of the primary harmonic *F* (figure 9(a)) of the spectrum and have selected the data points within 90% of this peak value. The data points within this 90% peak window, but belonging to the secondary peaks are then rejected and the remaining data points are used to plot the reduced PSD spectrum (shown by the shaded region in figure 9(b)). The integral of this plot gives an estimate of the signal power about the primary harmonic *F*. Here the signal power about the primary harmonic, $P_F$, is defined as $P_F = \ln(\text{integral of reduced PSD})$. For all the pulses we have obtained the $P_F$ parameter values and the histogram for gamma-ray and neutron induced events at 40.5°C and 44.5°C are shown in figure 10, where the $P_F$-axis is rescaled to zero corresponding to the lowest $P_F$ value. Figure 10 has similar characteristics as that of figure 7, implying that energy of the acoustic signal is mostly concentrated about the primary harmonic. Here also the gamma-ray induced events have lower $P_F$ values than the neutron induced events. As explained earlier this happens because for neutrons the nuclear recoils deposit more energy in the critical diameter compared to the lower LET electrons in the case of gamma-rays. More acoustic energy is released in the primary harmonic region of the signal produced during bubble nucleation induced by the higher LET nuclear recoils.

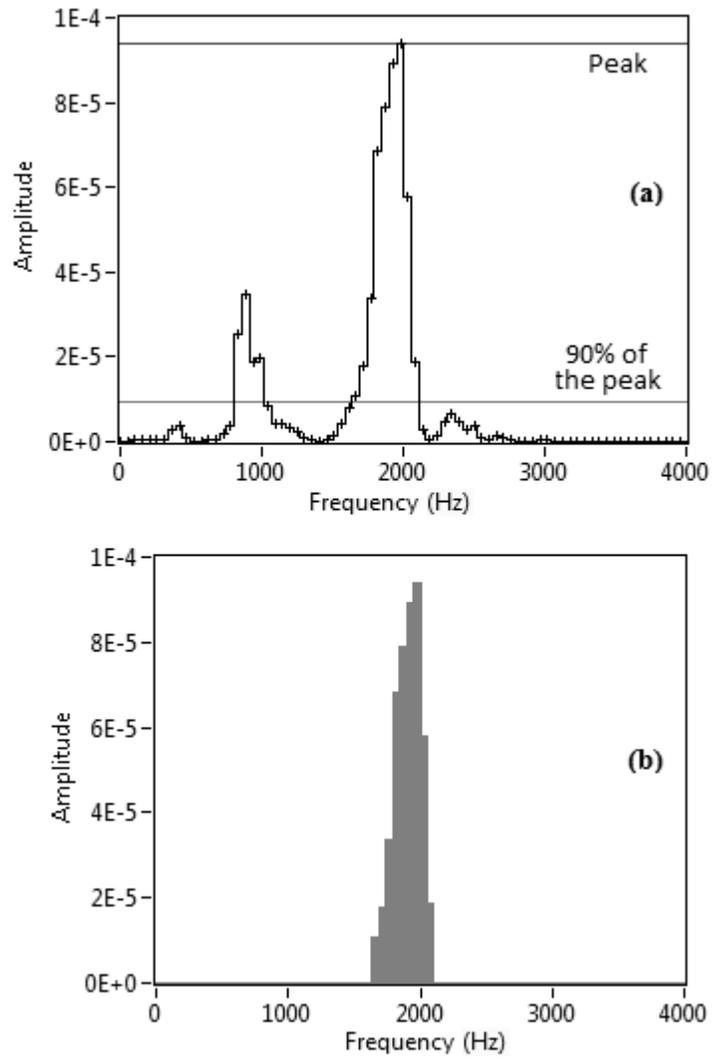

**Figure 9**. Typical power spectral density (PSD) of a pulse (a) and the reduced PSD obtained from it (b).

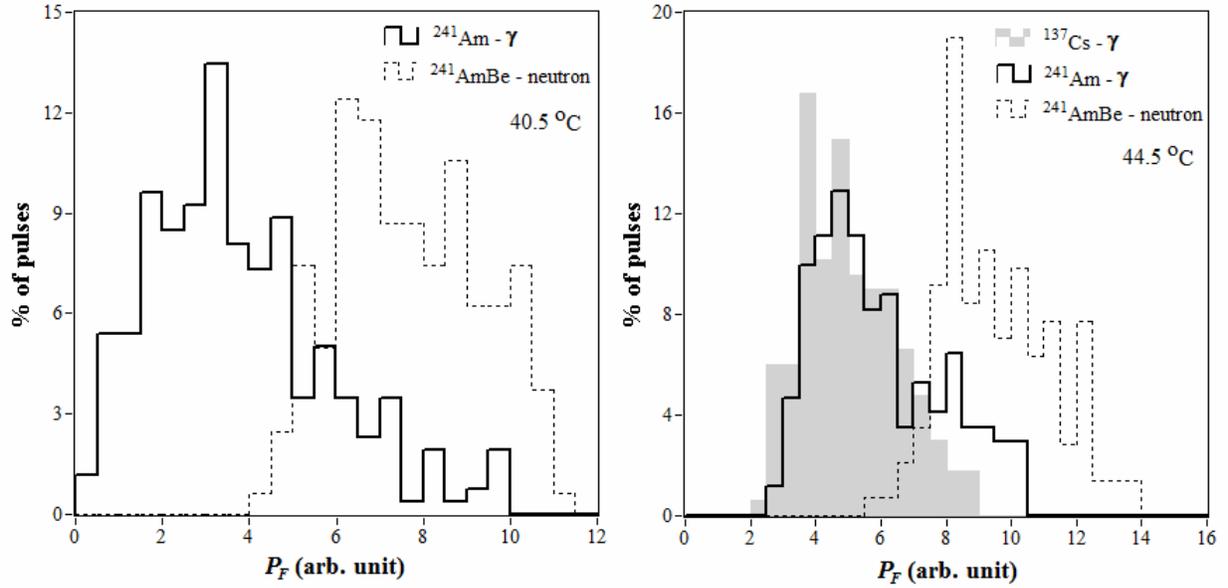

**Figure 10**. Distribution of signal power obtained from the reduced power spectral density plots of the neutron and gamma-ray induced events at 40.5°C and 44.5°C.

From the experimental results as shown in figures 6, 7, 8 and 10 it appears that for $^{241}$Am gamma-ray source the detector response shows a tail which overlaps with the respective neutron signals obtained with $^{241}$AmBe neutron source. Here the overlapping region for $^{137}$Cs and $^{241}$AmBe sources are much smaller as shown in figures 6, 7, 8 and 10 indicating the possibility of a neutron contamination in $^{241}$Am gamma-ray source. We have investigated the neutron contamination in the $^{241}$Am source by using the neutron monitor NM2B. The neutron dose rate at a distance of 11 cm from the $^{241}$Am source is about 6.5 µSv/h as measured using the NM2B neutron monitor. It is observed that the neutron dose rate decreases sharply with distance and at 40 cm it is about 1 µSv/h against the gamma-ray dose rate of about 200 µSv/h as measured using PM1402M radiation monitor.

## 5. Conclusion

The present work at low frequency demonstrates the variation of several parameters of the acoustic pulses for neutrons and gamma-rays induced events in superheated droplet detector. We observe a significant difference between the amplitudes, powers and frequencies of neutron and gamma-ray induced events in superheated R-12. Among the five parameters mentioned here, four parameters are useful in discrimination of neutron and gamma-ray induced events. This effect could be used to improve the gamma-ray background suppression in neutron detection with SDD in a mixed neutron gamma-ray radiation field.

**Acknowledgments**

The authors are grateful to Prof. B. K. Chatterjee, Department of Physics, Bose Institute, for providing the $^{241}$Am and $^{137}$Cs gamma sources and $^{241}$AmBe neutron source. The authors thank Mr. N. Biswas, Saha Institute of Nuclear Physics, for his assistance in the experiment.